\documentclass[twocolumn,showpacs,preprintnumbers,amsmath,amssymb,floatfix,nofootbib]{revtex4-1}
\usepackage{graphicx}
\usepackage{dcolumn}
\usepackage{bm}
\usepackage{color}
\usepackage{url}
\usepackage{amsmath}
\usepackage{hyperref}
\usepackage{subfigure}
\usepackage{cleveref}
\usepackage{enumitem}
\usepackage{rotating}
\usepackage{multirow}

\graphicspath{{./figures/}}
\newcommand{\be}{\begin{equation}}
\newcommand{\ee}{\end{equation}}
\newcommand{\ba}{\begin{eqnarray}}
\newcommand{\ea}{\end{eqnarray}}

\def\ltsima{$\; \buildrel < \over \sim \;$}
\def\simlt{\lower.5ex\hbox{\ltsima}}
\def\gtsima{$\; \buildrel > \over \sim \;$}
\def\simgt{\lower.5ex\hbox{\gtsima}}

\begin{document}
\title[EOS]{Demonstrating the feasibility of  probing the neutron star equation of state with second-generation gravitational wave detectors}

\author{Walter \surname{Del Pozzo}}
\email{walterdp@nikhef.nl}
\affiliation{Nikhef - National Institute for Subatomic Physics, Science Park 105, 1098 XG Amsterdam, The Netherlands}
\author{Tjonnie G.F. \surname{Li}}
\email{tgfli@nikhef.nl}
\affiliation{Nikhef - National Institute for Subatomic Physics, Science Park 105, 1098 XG Amsterdam, The Netherlands}
\author{Michalis \surname{Agathos}}
\email{magathos@nikhef.nl}
\affiliation{Nikhef - National Institute for Subatomic Physics, Science Park 105, 1098 XG Amsterdam, The Netherlands}
\author{Chris \surname{Van Den Broeck}}
\email{vdbroeck@nikhef.nl}
\affiliation{Nikhef - National Institute for Subatomic Physics, Science Park 105, 1098 XG Amsterdam, The Netherlands}
\author{Salvatore \surname{Vitale}}
\email{salvatore.vitale@ligo.mit.edu}
\affiliation{Massachusetts Institute of Technology, 185 Albany Street, Cambridge, MA 02139 USA}
\affiliation{Nikhef - National Institute for Subatomic Physics, Science Park 105, 1098 XG Amsterdam, The Netherlands}


\date{today}
\begin{abstract}
Fisher matrix and related studies have suggested that with second-generation gravitational wave detectors, it may be possible to infer the equation of state of neutron stars using tidal effects in binary inspiral. Here we present the first fully Bayesian investigation of this problem. 
We simulate a realistic data analysis setting by performing a series of numerical experiments of binary neutron star signals hidden in detector noise, assuming the projected final design sensitivity of the Advanced LIGO-Virgo network. With an astrophysical distribution of events (in particular, uniform in co-moving volume), we find that only a few tens of detections will be required to arrive at  strong constraints, even for some of the softest equations of state in the literature. Thus, direct gravitational wave detection will provide a unique probe of neutron star structure.
\end{abstract}


\maketitle

\emph{Introduction.} The Advanced LIGO \citep{advligo} and Virgo \citep{advvirgo} gravitational wave (GW) detectors are expected to start taking data in 2015, with gradual upgrades in the following years. KAGRA \citep{lcgt} in Japan and possibly LIGO-India \cite{indigo} will come online few years later. Second-generation instruments may detect tens of GW signals from compact binary coalescences: the rates are expected to be in the range $\sim 1-100$ yr$^{-1}$ conditional on the astrophysical event rate, the instruments' duty cycles, and the sensitivity evolution of the detectors \citep{cbc-low-mass-S5VSR1}. 

Currently, predictions for the neutron star equation of state (EOS) vary by an order of magnitude in terms of tidal deformability \citep{Hinderer2010}. The detection of gravitational wave signals from coalescing binary neutron stars (BNS), or a neutron star and a black hole (NSBH), could provide the missing information. During the last stages of inspiral, the Newtonian tidal field $\mathcal{E}_{ij}$ of one component will induce a quadrupole moment $Q_{ij}$ in the other, where to leading order in the adiabatic approximation $Q_{ij} = -\lambda(\mbox{EOS}; m)\,\mathcal{E}_{ij}$. The tidal deformability parameter $\lambda(\mbox{EOS}; m)$ depends on the neutron star mass $m$ in a way that is determined by its EOS. The neutron stars' deformation has an influence on the orbital motion, in particular the phase, which up to a factor of two is also the phase $\Phi(t)$ of the emitted gravitational wave signal. In the post-Newtonian approximation one has $\Phi(t) = \Phi_{\rm PP}(t) + \Phi_{\rm tidal}(t)$, where $\Phi_{\rm PP}$ is the phase for point particles, and the tidal contribution $\Phi_{\rm tidal}$ takes the form \cite{Hinderer2010,Vines2010}
\begin{widetext}
\begin{equation}
\Phi_{\rm tidal}(v) 
= \sum_{a=1}^2 \frac{3\lambda_a}{128 \eta M^5}\left[ -\frac{24}{\chi_a} \left(1 + \frac{11 \eta}{\chi_a}\right)\,\left(\frac{v}{c}\right)^5 -\frac{5}{28 \chi_a} \left(3179 - 919\,\chi_a - 2286\,\chi_a^2 + 260\,\chi_a^3 \right)\,\left(\frac{v}{c}\right)^7
\right],
\label{Phitidal}
\end{equation} 
\end{widetext}
where the sum is over the components of the binary,  $v = (M\omega)^{1/3}$ is a characteristic velocity in terms of the gravitational wave frequency $\omega$, $\chi_a = m_a/M$, $\lambda_a = \lambda(m_a)$ where $m_a$ are the component masses, $M$ is the total mass, and $\eta = m_1 m_2/M^2$. The function $\lambda(m)$ takes the form $\lambda(m) = (2/3)\,k_2\,R^5(m)$, with $k_2$ the second Love number and $R(m)$ a neutron star's radius as a function of mass. Note that $\lambda(m)$ enters Eq.~(\ref{Phitidal}) only in the combination $\lambda(m)/M^5 \propto (R/M)^5\sim 10^2 - 10^5$ \cite{Lattimer2012}. Hence, although tidal effects only enter at very high post-Newtonian order (5PN and 6PN, in the usual notation), they come with a large prefactor, so that they might be observable even with second-generation detectors. 

Read \emph{et al.}~\cite{ReadEtAl:2009} estimated that a single detection of a close-by BNS source (100 Mpc) could constrain the neutron star radius to 10\%. 
Hinderer \emph{et al.}~\cite{Hinderer2010} performed a Fisher matrix calculation with post-Newtonian waveforms truncated at 450 Hz to see how well $\lambda$ might be measurable for close-by BNS from the low-frequency inspiral part alone. Their results suggest that even for a very hard EOS, corresponding to the largest tidal deformability, it would be difficult to extract information about the EOS from this frequency regime with the upcoming second-generation detectors. Damour, Nagar, and Villain \cite{Damour2012} assumed an approximation to effective one-body waveforms, which they used to the point where the neutron stars are touching. Their Fisher matrix analysis indicated more encouraging prospects, suggesting that it might be possible after all to gain information about the EOS. Lackey \emph{et al.} \cite{Lackey2012} performed similar analyses for NSBH but using hybrid numerical relativity waveforms matched to effective one-body approximants, also arriving at cautiously optimistic conclusions. The abovementioned studies were for single detected sources; a first investigation for multiple sources was reported in \cite{MarkakisEtAl:2010}, where it was estimated that a similar accuracy as in \cite{ReadEtAl:2009} could be achieved with 3 low signal-to-noise ratio (SNR) detections. On the other hand, Fisher matrix based analyses are known to be unreliable at low SNR \cite{Cokelaer2008,Vallisneri2008,Zanolin2010}, and from a realistic data analysis perspective, these studies still leave unclear that it will be possible to make strong statements about the EOS even with multiple sources. 

We present the first Bayesian investigation of the problem, in a realistic data analysis setting. In particular, we consider BNS signals in simulated detector noise, assuming the projected final design sensitivity of the Advanced LIGO-Virgo network. Sources are distributed in an astrophysically realistic way. We evaluate two different Bayesian methods which allow us to combine information from multiple sources. We find that a few tens of sources will be required to arrive at strong constraints, even for some of the softest equations of state in the literature. Thus, direct gravitational wave detection will provide a unique probe of neutron star structure.\\ 

\noindent
\emph{Assumptions.} At the time this work was started, the waveform model of \cite{Damour2012}, which was inspired on the effective one-body formalism and has tidal terms to higher PN order, was not yet available. We consider the post-Newtonian frequency domain approximant of \cite{Hinderer2010} with tidal contributions at 5PN and 6PN. We cut this off at the ``last stable orbit" (LSO) frequency $f_{\rm LSO} = 1/(6^{3/2} \pi\,M)$. Since spins are expected to be small in binary neutron star systems \cite{OShaughnessy:2008}, we neglect them. Our waveform model also suffers from the absence in the phase of unknown point particle contributions beyond 3.5PN. 
These will be set by the neutron star masses (and spins, but their effects will be minor), knowledge of which is obtained primarily from the low frequency regime (to a fraction of a percent for ``chirp mass" $\mathcal{M} = M\eta^{3/5}$ and about a percent for $\eta$) whereas tidal effects are measured from the high frequency part of the waveform \cite{Damour2012}. For this reason, when the coefficients of the unknown 4PN-6PN contributions become available, we do not expect them to act as nuisance in inferring the EOS. 
Overall it seems reasonable to assume that results obtained with our waveform model will be indicative of what can be achieved with second-generation detectors.   

Redshift effects are included in $\Phi_{\rm PP}(v)$, assuming a $\Lambda$CDM cosmology with $H_0 = 70\,\mbox{km}\,\mbox{s}^{-1}\,\mbox{Mpc}^{-1}$. Instead of using the Fisher matrix formalism, we perform full Bayesian analyses on signals that are coherently added to simulated stationary, Gaussian noise following the predicted Advanced LIGO and Virgo final design sensitivities. The BNS sources have component masses that are drawn uniformly from the interval $[1, 2]\,M_\odot$. Their sky positions, inclinations and polarizations are uniform on the sphere. Sources are distributed uniformly in co-moving volume, with luminosity distances between 100 and 250 Mpc, so that the majority of events will be near the threshold of detectability, chosen at a network SNR of $8$ \cite{Cutler1993}; this means that 70\% will be at a distance greater than 175 Mpc, and only 5\% will be closer than 120 Mpc.\\

\noindent
\emph{Method 1: Taylor expansion of $\lambda(m)$.} One way to obtain information about the EOS is by expanding the tidal deformability in $(m-m_0)/M_\odot$, with $m_0$ some reference mass:
\begin{equation}
\lambda(m) = \sum_j \frac{1}{j!}\,\lambda_j\,\left(\frac{m - m_0}{M_\odot}\right)^j.
\label{expansion}
\end{equation}
For a given EOS, the coefficients $\lambda_j$ are fixed. This provides us with a way to combine information about the EOS from multiple sources. Let $d_1, d_2, \ldots, d_N$ be $N$ stretches of 3-detector data, each containing a detected BNS signal, and denote whatever additional information we hold by $I$. Assuming that all systems have the same EOS, the posterior density functions $p(\lambda_j | d_n, I)$ from each of the detections $d_n$ together yield a combined posterior density
\begin{equation}
p(\lambda_j| d_1, d_2, \ldots, d_N, I) = p(\lambda_j|I)^{1-N} \prod_{n=1}^N p(\lambda_j|d_n, I),
\label{combinedpdf}
\end{equation}
where we have assumed independence of the $d_n$ and used Bayes' theorem; $p(\lambda_j|I)$ is the prior density for the parameter $\lambda_j$. 

Only a limited number of coefficients in Eq.~(\ref{expansion}) will be measurable. Moreover, if too many coefficients are estimated at once, the measurement accuracy on all of them will deteriorate. In practice, already $\lambda_2$ can not be measured. Therefore, in the recovery waveforms, we adopt a linear approximation of  $\lambda(m)$ around the ``canonical" reference mass $m_0 = 1.4\,M_\odot$ \cite{Damour2012}:
\begin{equation}
\lambda(m) \simeq \lambda_0 + \lambda_1\,(m - 1.4\,M_\odot)/M_\odot.
\label{linearized}
\end{equation}
Now, for each detection $d_n$ we need to compute the posterior probability densities $p(\lambda_0|d_n,I)$ and $p(\lambda_1|d_n, I)$. These are obtained by marginalizing over all the other parameters in the problem; for instance,
\begin{equation}
p(\lambda_0 | d_n, I) = \int\, d\vec{\theta}\,d\lambda_1\,p(\vec{\theta}, \lambda_0, \lambda_1| d_n, I),
\end{equation}
where $\vec{\theta}$ represents masses, sky position, orientation of the orbital plane, and distance. The joint posterior density function for \emph{all} the parameters takes the form
\begin{equation}
p(\vec{\theta}, \lambda_0, \lambda_1| d_n, I) = \frac{p(d_n |\vec{\theta}, \lambda_0, \lambda_1, I)\,p(\vec{\theta}, \lambda_0, \lambda_1| I)}{p(d_n | I)}.
\label{jointposterior}
\end{equation}
Here $p(\vec{\theta}, \lambda_0, \lambda_1| I) = p(\vec{\theta} | I)\,p(\lambda_0 | I)\,p(\lambda_1 | I)$. The prior density $p(\vec{\theta} | I)$ is taken to be the same as in \cite{Li2012a}. We express $\lambda(m)$ in units of $\mbox{s}^5$. For $p(\lambda_0 | I)$ we choose a flat distribution in the range $[0,5] \times 10^{-23} \,\mbox{s}^5$, and for $p(\lambda_1 | I)$ a flat distribution on $[-5,0]\times 10^{-18} \,\mbox{s}^4\,M_\odot$; these choices cover all the EOS considered in \cite{Hinderer2010}. The prior probability for the data, $p(d_n|I)$, is obtained by demanding that the left hand side of (\ref{jointposterior}) be normalized. Finally, the \emph{likelihood} is given by \cite{Veitch2010}   
\begin{eqnarray}
&& p(d_n | \vec{\theta}, \lambda_0, \lambda_1, I)  \nonumber\\
&& = \mathcal{N}\,\exp\left[- 2  \int_{f_0}^{f_{\rm LSO}} df \,\frac{|\tilde{d}_n(f) - \tilde{h}_{\rm lin}(\vec{\theta},  \lambda_0, \lambda_1; f)|^2}{S_n(f)}\right], 
\end{eqnarray}
where $\mathcal{N}$ is a normalization factor, $\tilde{d}_n$ is the Fourier transform of the data stream for the $n$th detection, and $S_n(f)$ is the one-sided noise power spectral density; $f_0$ is a lower cut-off frequency, which we take to be 20 Hz. $\tilde{h}_{\rm lin}(\vec{\theta},  \lambda_0, \lambda_1; f)$ is our frequency domain waveform, with the linearized expression for $\lambda(m)$, Eq.~(\ref{linearized}), substituted into the tidal contribution to the phase, Eq.~(\ref{Phitidal}). To explore the likelihood function, we used the method of Nested Sampling as implemented by Veitch and Vecchio \cite{Veitch2010}. 

\begin{figure}[!h]
\includegraphics[width=3.0in]{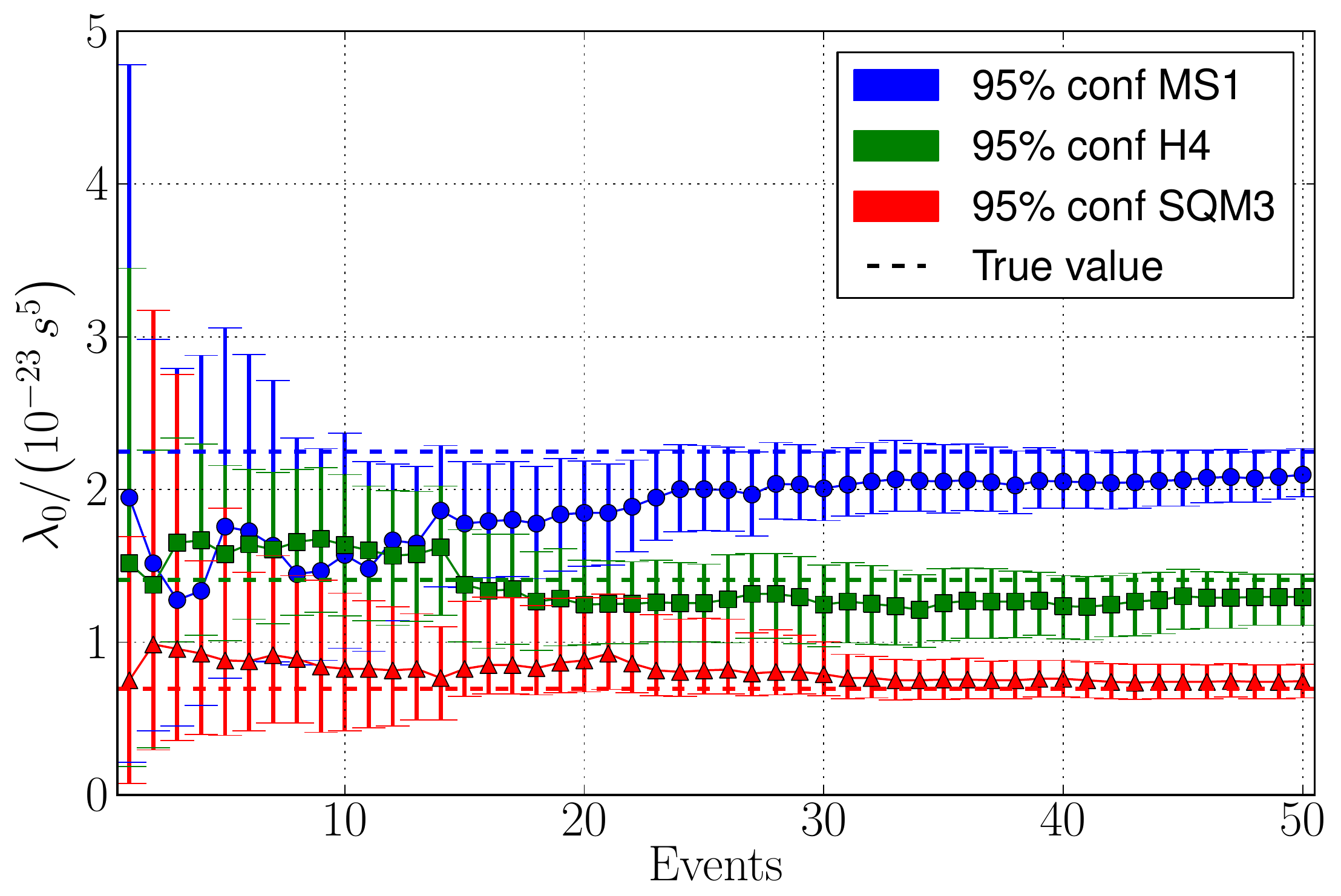}
\caption{Median and 95\% confidence interval evolution for the $\lambda_0$ parameter as an increasing number of sources is taken into consideration, for three different equations of state in the signals: a hard (MS1), a moderate (H4), and a soft (SQM3) EOS. In each case, the dashed line indicates the true value.}
\label{f:l0}
\end{figure}

In Fig.~\ref{f:l0}, we show the evolution with an increasing number of sources of the medians and 95\% confidence intervals in the measurement of $\lambda_0$, for three different EOS models from Hinderer \emph{et al.}~\cite{Hinderer2010}: a hard EOS (MS1), a moderate one  (H4), and a soft one (SQM3). In each case, after a few tens of sources, the value of $\lambda_0$ is recovered with a statistical uncertainty $\sim 10\%$, and it is easily distinguishable from the ones for the other EOS. (On the other hand, $\lambda_1$ remains uncertain.) We see that the posterior medians for $\lambda_0$ are ordered correctly, which suggests a second method to identify the EOS, namely hypothesis ranking.\\ 

\begin{figure*}[!ht]
\includegraphics[width=7in]{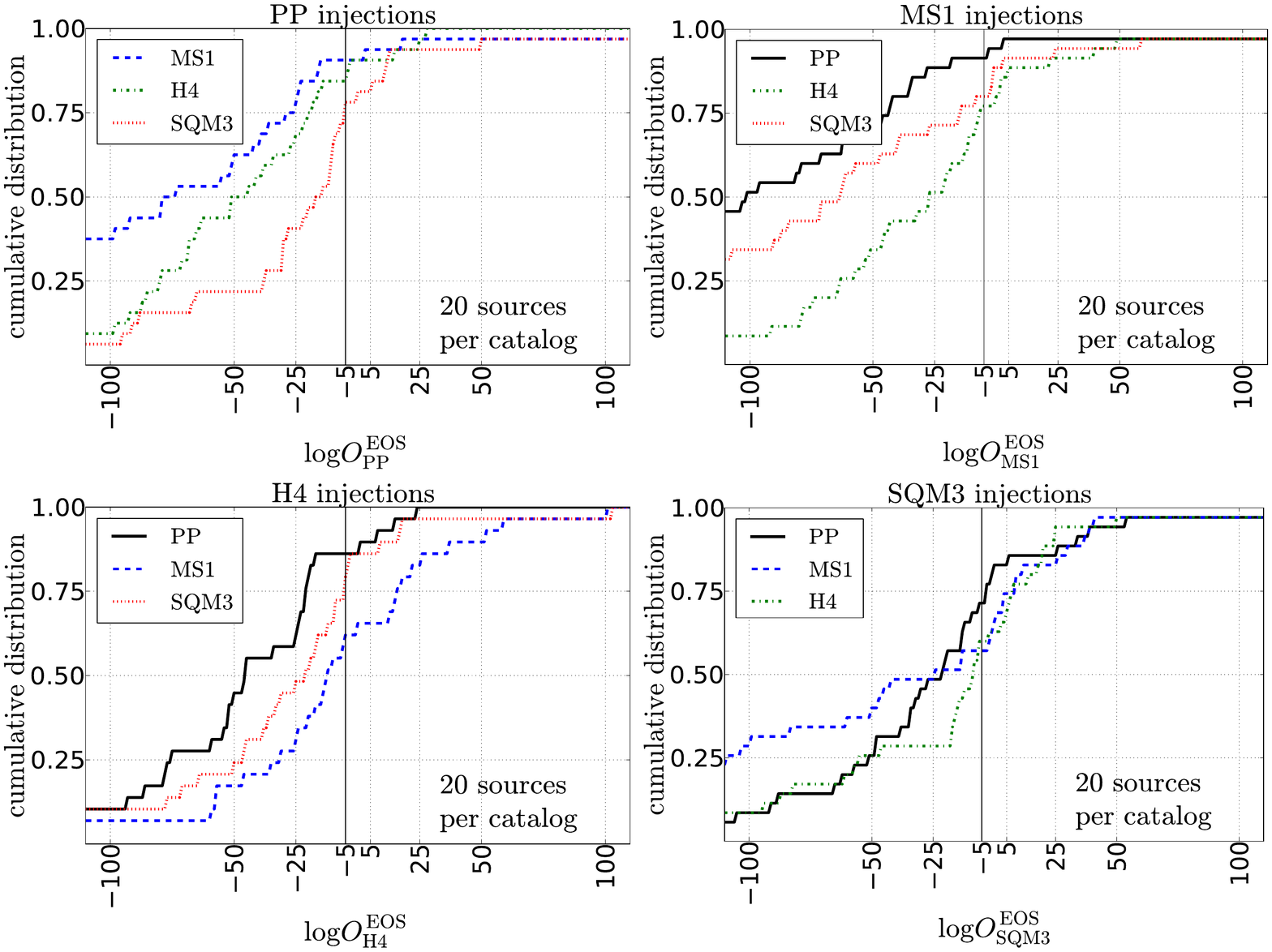}
\caption{Cumulative distributions of log odds ratios for $\mathcal{O}(30)$ simulated catalogs of sources, for various EOS against the true EOS model, with different panels corresponding to different true EOS in the signal model (stated at the top of each panel). For each signal model, the number of sources per catalog was fixed to 20. Using the Jeffreys criterion, we consider the true EOS, $\mathcal{H}_j$, to be identified correctly if $\ln O^i_j < -5$ (odds less than 1:150) for all the other EOS, $\mathcal{H}_i$. In the cases where the true EOS are MS1, H4, and SQM3, this happens for 77\%, 62\%, and 57\% of catalogs, respectively. Note how hypotheses tend to get ranked correctly by ``goodness of fit", \emph{i.e.}~hardness of the EOS.\label{f:cum-dist}}
\end{figure*}

\noindent
\emph{Method 2: Hypothesis ranking.} Hinderer \emph{et al.}~computed the function $\lambda(m)$ for a large number of (families of) equations of state, some of them mainly involving neutrons, protons, electrons, and muons, others allowing for pions and hyperons, and a few assuming strange quark matter. Given a (arbitrarily large) discrete set $\{\mathcal{H}_k\}$ of models, each corresponding to a different EOS, or equivalently a different deformability $\lambda(m)$, the relative \emph{odds ratios} for any pair of models $\mathcal{H}_i$, $\mathcal{H}_j$ can be computed as
\begin{equation}
O^i_j = \frac{P(\mathcal{H}_i | d_1, d_2, \ldots, d_N, I)}{P(\mathcal{H}_j | d_1, d_2, \ldots, d_N, I)}.
\end{equation}
Again assuming independence of the detector outputs $d_1, d_2, \ldots, d_N$ and using Bayes' theorem, one can write
\begin{equation}
O^i_j = \frac{P(\mathcal{H}_i | I)}{P(\mathcal{H}_j | I)} \prod_{n = 1}^N \frac{P(d_n | \mathcal{H}_i, I)}{P(d_n |\mathcal{H}_j, I)}.
\end{equation}
$P(\mathcal{H}_i | I)$ is the probability of the model $\mathcal{H}_i$ before any measurement has taken place, and similarly for $\mathcal{H}_j$; in the absence of more information,  these can be set equal to each other for all models $\mathcal{H}_k$. The evidences for the various models are given by
\begin{equation}
p(d_n | \mathcal{H}_k, I) = \int d\vec{\theta}\,p(d_n | \mathcal{H}_k, \vec{\theta}, I)\,p(\vec{\theta} | I),
\end{equation}
with $\vec{\theta}$ the parameters of the template waveforms (masses, sky position, etc.) and $p(\vec{\theta} | I)$ the prior probabilities for these parameters, which we choose to be the same as in \cite{Li2012a}. The likelihood function $p(d_n | \mathcal{H}_k, \vec{\theta}, I)$ takes the form
\begin{eqnarray}
&&p(d_n | \mathcal{H}_k, \vec{\theta}, I) \nonumber\\ 
&&= \mathcal{N}\,\exp\left[- 2  \int_{f_0}^{f_{\rm LSO}} df \,\frac{|\tilde{d}_n - \tilde{h}_k(\vec{\theta}; f)|^2}{S_n(f)}\right]. 
\end{eqnarray}
This time $\tilde{h}_k(\vec{\theta}; f)$ is the waveform model corresponding to the EOS $\mathcal{H}_k$, meaning the abovementioned frequency domain approximant with tidal contributions to the phase as in Eq.~(\ref{Phitidal}), with a deformability $\lambda(m)$ corresponding to that EOS. Here too, we use Nested Sampling to probe the likelihood \cite{Veitch2010}. 

The set $\{\mathcal{H}_k\}$ could comprise all the models considered in \emph{e.g.}~\cite{Hinderer2010}, and many more. In this Letter we wish to show that it will at least be possible to distinguish between a hard, a moderate, and a soft EOS. Accordingly, we focus on just three EOS models, the ones labeled MS1, H4, and SQM3 in \cite{Hinderer2010}. In addition we consider the point particle model (PP) in which $\lambda(m) \equiv 0$. 
Fig.~\ref{f:cum-dist} shows the cumulative distribution of $\ln O^k_j$ for different signal models $\mathcal{H}_k$ against the true EOS model $\mathcal{H}_j$, for $\mathcal{O}(30)$ simulated catalogs of 20 sources each. A useful criterion for correct identification of the underlying EOS is that the log odds ratio of the incorrect models against the true EOS be \emph{decisive} according to the Jeffreys scale, \emph{i.e.}~$< - 5$ in log odds (odds less than 1:150, which one can think of as being roughly similar to 3 $\sigma$) \cite{Jeffreys}. When the signals' EOS is MS1 (top right panel of Fig.~\ref{f:cum-dist}), we see that the runner-up model, H4, is decisively disfavored ($\ln O^{\rm H4}_{\rm MS1} < -5$) for over 77\% of simulated catalogs of sources. When the true EOS is H4 (bottom left panel of the Figure), the runner-up is SQM3, which is decisively disfavored ($\ln O^{\rm SQM3}_{\rm H4} < -5$) for 62\% of catalogs. Finally, when the underlying EOS is SQM3 (bottom right panel), correct identification of the EOS happens for about 57\% of catalogs. (To give an indication of what happens with a smaller number of detections: for 10 sources per catalog, these fractions would have been 68\%, 46\%, and 49\%, respectively.) We stress that Fig.~\ref{f:cum-dist} pertains to only 20 detected sources; the results will improve as more detections are made.


Fig.~\ref{f:cum-dist} also shows another interesting feature of the approach presented in this section: the odds ratio ranks the various competing hypotheses according to their ``goodness of fit''. For example, the top left panel shows the odds ratio for catalogs of 20 sources for PP signals. All finite size models are correctly disfavored compared to the PP hypothesis, and the degree of belief in the three competing models reflects the size of the physical effects they predict: the harder the EOS the less we should be inclined to believe that it faithfully describes our observations. 
This feature suggests that in a real GW detection scenario, even if none of the EOS models considered will be the one chosen by Nature, we will still be able to rank the models according to the predictions they offer and thus guide the development of theoretical models for the interiors of neutron stars.\\ 

\noindent
\emph{Conclusions and future work.} We have shown that, in a realistic data analysis setting, (a)  quantitative information about the size of the tidal deformability at a given reference mass can be obtained with a 2-$\sigma$ statistical uncertainty of $\sim 10\%$ after a few tens of detections, and (b) hypothesis ranking will be able to distinguish between a hard, moderate, and soft EOS with $\mathcal{O}(20)$ events.

Our results open the door for further studies. 
For example, can we arrive at more direct physical information about pressure as a function of density? Already in 1992, Lindblom noted that measurements along the neutron star mass-radius curve can be converted to points along the pressure versus density curve \cite{Lindblom1992}. Other possibilities include probing pressure against density represented as piecewise polytropes \cite{ReadEtAl:2009} or using spectral fits \cite{Lindblom2010}. Also, in the preliminary study presented here, inspiral waveforms were terminated at the LSO for the point particle limit, but depending on the EOS, LSO could occur earlier than that \cite{Bejger2005}, which in itself is a source of information. Moreover, as shown in \cite{ReadEtAl:2009}, for hard equations of state, the formation of a black hole could be preceded by the occurrence of a fast-rotating, highly asymmetric hypermassive neutron star, leading to a distinctive post-merger signal which may be detectable with advanced gravitational wave observatories. In this regard we also note the equation-of-state dependence of the post-merger phase found in \cite{Bauswein2012}. Extracting this information will most likely require the construction of phenomenological waveforms which have a close match to numerical ones; see \emph{e.g.}~\cite{Lackey2013}. Finally, in the inspiral we neglected the effects of spins; although these are expected to be very small for binary neutron stars and are unlikely to significantly affect our conclusions, they should nevertheless be quantified as well.    
Ultimately, our findings motivate the construction of a full data analysis pipeline to constrain the EOS of neutron stars using BNS detections.\\ 



\noindent
\emph{Acknowledgements.} WDP, TGFL, MA and CVDB are supported by the research programme of the Foundation for Fundamental Research on Matter (FOM), which is partially supported by the Netherlands Organisation for Scientific Research (NWO). SV acknowledges the support of the National Science Foundation and the LIGO Laboratory. LIGO was constructed by the California Institute of Technology
and Massachusetts Institute of Technology with funding from the National Science Foundation and operates under cooperative agreement PHY-0757058. The authors would like to acknowledge the LIGO Data Grid clusters, without which the simulations could not have been performed. Specifically, these include the computing resources supported by National Science Foundation awards PHY-0923409 and PHY-0600953 to UW-Milwaukee. Also, we thank the Albert Einstein Institute in Hannover, supported by the Max-Planck-Gesellschaft, for use of the Atlas high-performance computing cluster. The authors wish to acknowledge M.C.~Miller, J.S.~Read, and R.~Sturani for useful comments and discussions.

%
%

\end{document}